# $Gd_{0.2}Ce_{0.8}O_{1.9}/Y_{0.16}Zr_{0.84}O_{1.92}$ nanocomposite thin films for low temperature ionic conductivity

Giovanni Perin,[1,2,*] Christophe Gadea,[1] Massimo Rosa,[1] Simone Sanna,[1] Yu Xu,[1] Ragnar Kiebach,[1] Antonella Glisenti,[2] and Vincenzo Esposito [1,*].

[1]: Technical University of Denmark, Department of Energy Conversion and Storage, Frederiksborgvej 399, 4000, Roskilde, Denmark.

[2]: University of Padova, Department of Chemical Sciences, via Marzolo 1, 35131, Padova, Italy.

*: Corresponding authors:

Giovanni Perin

Dipartimento di Scienze Chimiche, Università di Padova

Via F. Marzolo, 1 - 35131 Padova, Italy

Ph. +39 049 8275858 - E-mail: giovanni.perin.1@phd.unipd.it

Prof. Vincenzo Esposito

Technical University of Denmark, Department of Energy Conversion and Storage

Frederiksborgvej 399 - 4000, Roskilde, Denmark

Ph. +45 46 77 56 37 - E-mail: vies@dtu.dk

**Abstract:** $Gd_{0.2}Ce_{0.8}O_{1.9}/Y_{0.16}Zr_{0.84}O_{1.92}$ (GDC/YSZ) nanocomposite is synthesized by a novel hybrid chemical route, where colloidal crystalline GDC nanoparticles from continuous hydrothermal synthesis are dispersed into a metalorganic YSZ matrix precursor. The result is a mixture of metal oxides in which GDC nanoparticles are finely distributed in a continuous metalorganic polymeric matrix to be crystallized after calcination. The GDC nanoparticles reduce the temperature necessary to obtain crystalline YSZ, which is already formed at 400 °C. The nanocomposite reveals structural stability up to 800 °C when treated in both air and reducing atmosphere, showing the onset of diffusion below 1000 °C. The diffusional processes are largely dependent on the nanometric grain size, with $Zr^{4+}$ diffusing abruptly towards GDC in air at 1000 °C and GDC/YSZ interdiffusion being hindered in reducing environment despite the onset temperature of 900 °C.

The nanocomposite precursor is an inkjet-printable reactive water-based material, suitable for the deposition of thin films with a thickness below 100 nm after calcination at 750 °C. The crystal structure of the film reveals no interaction between GDC and YSZ but a microstrain (0.3% tensile strain for YSZ). The thin film microstructure shows a compact layer with 94% density.

The nanocomposite shows high oxygen ionic conductivity at low temperatures (> $5 \cdot 10^{-3}$ S·cm$^{-1}$ at 500 °C), low activation energy (0.55 eV), and dominant oxygen ionic conductivity even in reducing conditions (pO$_2$ < $10^{-25}$ atm). We show that these properties arise from the large interface between the components of the composite, due to the embedding of the GDC nanoparticles in the YSZ matrix, while ZrO-CeO intermixing can be avoided and no n-type conductivity is observed even at low oxygen activities and high temperatures.

## 1. Introduction

Solid state ionic conductors, i.e. solid state ionics, are an important family of materials that can transport electrical current in through ions, either positive (e.g. Li$^+$, H$^+$, Na$^+$) or negative (O$^{2-}$, OH$^-$). They are the building blocks of numerous energy conversion and storage devices, including batteries, fuel cells, membranes and sensors.

Many of these devices are based on oxygen deficient metal oxides, which are used as O$^{2-}$-ions electrolytes and electrodes in the form of ceramic materials. In the case of electrolytes, a dense layer ensures separation and gas-tightness between two chemical environments of different reactive atmospheres. Electrolytes also require nearly pure ionic transport in order to avoid electronic losses and drops of efficiency[1–3]. For electrodes and membranes mixed ionic-electronic conductivity (MIEC) is required; often this result is achieved by mixing electrochemically/catalytically active and electronic conducting materials with ionic conductors, especially with the aim to enhance contact zone suitable for solid-gas conversion reactions[4,5].

Among the O$^{2-}$-conductors, defective oxides with AO$_2$ fluorite structure have been intensively investigated due to their high ionic conductivity and chemical stability, with yttrium stabilized zirconium oxide (YSZ) and gadolinium doped ceria (GDC) still representing the state of art materials[6,7]. As for zirconia, the highest performing phase is the cubic structure which is stabilized by trivalent cations[8], *e.g.* Y$^{3+}$, that introduce extrinsic oxygen vacancies, resulting in a material with wide thermochemical stability and nearly pure ionic conductivity of ~$10^{-2}$ S·cm$^{-1}$ at 800 °C in both low and high $O_2$ gas concentrations[9]. As opposed to zirconia, cerium oxide ($CeO_2$, ceria) naturally occurs as cubic fluorite, having a high concentration of intrinsic oxygen defects due to the multivacancy of $Ce^{3+/4+}$. Extrinsic oxygen vacancies concentration can be further enhanced *via* acceptor dopants, *e.g.* by $Gd^{3+}$, reaching a typical ionic conductivity of ~$10^{-2}$ S·cm$^{-1}$ at just 600 °C[9]. Despite being more conductive than YSZ, at low oxygen activity and high temperature (T > 600 °C, $pO_2$ < $10^{-20}$ atm) doped-ceria exhibits a non-negligible electronic conduction due to small-polarons arising from the co-existence of $Ce^{4+}/Ce^{3+}$[10]. $Ce^{4+}$ reduction to $Ce^{3+}$ is also accompanied by a severe increase (up to 2%) in the crystalline cell's volume and fast interdiffusive effects at low temperatures[11]. The choice for this material is thus limited to the working conditions that allow to prevent uncontrolled expansion, interdiffusion and consequent degradation phenomena occurring during processing and operation[12–15].

A strategy for avoiding the drawbacks of diffusion consists of lowering the working temperatures. In order to preserve the performances the thickness of the layers is reduced, *i.e.* working with materials in the form of thin films[16–19]; this solution is widely adopted in miniaturized electrochemical energy devices for low temperatures[20]. Nano-designing allows coupling different ionic conductors in multilayers, composites and heterostructures, where the role of the interface becomes more relevant than the properties of the constitutive materials[21–24]. As already proved by multi-layered electrolytes, merging the YSZ stability at low $pO_2$ with the high ionic conduction of GDC is appealing, but could be challenging, especially in traditional ceramic processing routes[25]. In fact GDC and YSZ have the same crystal structure and are reciprocally soluble in a wide range of compositions, and during co-sintering can form ceria-zirconia solid solutions that are generally characterized by a lower ionic conductivity than the parent structures[26–28].

Elemental diffusion, and thus mass diffusion, in defective oxides is a complex effect that is controlled by oxygen defects and cation migration, which can be dramatically changed with the calcination atmosphere[29]. As an example, previous investigations showed that in air $Zr^{4+}$ is the prevalent specie diffusing into the GDC lattice, while in reducing conditions the diffusion of $Ce^{3+}$ is predominant, due to larger concentration of oxygen defects [30,31]. Moreover, diffusion is also affected by the presence and the amount of doping elements or by the microstructure of the sample, since lattice and grain boundary processes are characterized by different energies[32,33]. Therefore, consolidation and crystallization of GDC/YSZ composites has to be designed carefully to occur at low temperatures and avoiding the formation of the undesired phases, whose effect can be even more dramatic at the nanoscale; in fact, where a large interface between the components exists, fast diffusion mechanisms can be easily unleashed by unavoidable thermochemical treatments[34–36].

Among the numerous technologies for the deposition of thin films a low temperature, inkjet printing is growing in consideration because of its appealing advantages such as low cost, use in 3D printing platforms, possibility of continuous and customizable printing, no need of vacuum conditions for the deposition and large availability of printable materials and substrates. For the ink formulations, these are generally either colloidal inks, based on powders, *e.g.* nanoparticles, or reactive metalorganic precursors [37,38]. A main advantage of this technique consists of the possibility of reducing the processing temperatures necessary for the consolidation of the ceramic layers, which are typically lower than the ones used for traditional sintering treatments[39].

In this work, we present a novel strategy developed for obtaining a dense nanocomposite combining the advantages of GDC and YSZ. The material is tailored to be inkjet-printable for the realization of thin film ionic conductors. To gain control on the stoichiometry, a hybrid chemical strategy is adopted: pre-formed GDC nanoparticles are finely dispersed in a reactive metalorganic sol-gel solution used as precursor of YSZ. The GDC nanoparticles here are adopted to enhance the electrical properties, while the sol-gel is used to form a dense YSZ matrix at low temperature.

Due to both the critical instability of GDC in reducing conditions and the possible formation of detrimental $CeO_2$-$ZrO_2$ solid solution, the evolution of crystal structure, composition and microstructure of the nanocomposite material is investigated. A thin film is then produced by inkjet printing and the electrical properties are characterized and compared to the properties of bulk and nanocrystalline GDC and YSZ, and to those of other nanocomposites.

## 2. Experimental

### 2.1 Hybrid $Gd_{0.2}Ce_{0.8}O_{1.9}$/$Y_{0.16}Zr_{0.84}O_{1.92}$ synthesis

The nanocomposite was synthesized by means of dispersing GDC nanoparticles into a reactive YSZ sol-gel matrix.

The colloidal $Gd_{0.2}Ce_{0.8}O_{1.9}$ nanoparticles were synthesized by continuous hydrothermal flow synthesis in the form of water based colloids, as also reported in a previous work[40]. The nanoparticles were subsequently dispersed adding Dispex A40 (Ciba-BASF, UK) to the aqueous solution using an ultrasonic processor (Hielscher UP200St, Germany). The solid loading of the Gd$_{0.2}$Ce$_{0.8}$O$_{1.9}$ nanoparticles solution was determined to be 7.1% wt/wt by thermogravimetric analysis (TGA, TG 209 F1 Libra, Netzsch, Germany).

The 8%Y$_2$O$_3$-ZrO$_2$ sol-gel reactive ink was synthesized using N-methyl-diethanolamine (MDEA) ligand to chelate the zirconium precursor (Zirconium(IV)-propyl solution, 70% wt in 1-propanol, Sigma-Aldrich) to avoid uncontrolled reaction. A complexation ratio (r=[MDEA]/[Zr(OPr)$_4$]) of 12 was chosen, as it previously proved to have good printability and long term stability[38,41]. The preparation of the ink (schematized in **Fig. 1a**) consisted of stabilization in argon atmosphere of the Zr-propoxide with MDEA, to prevent hydrolization and gelification of the precursor, followed by the addition of the colloidal Gd$_{0.2}$Ce$_{0.8}$O$_{1.9}$ nanoparticles dispersed in aqueous solution. The Gd$_{0.2}$Ce$_{0.8}$O$_{1.9}$/Y$_{0.16}$Zr$_{0.84}$O$_{1.92}$ molar ratio was fixed to 1:3.3 in order to prevent clogging of the nozzles during printing[37].

The obtained ink was calcined for 1h in synthetic air (flow rate 25 sccm) at different temperatures from 400 to 1400 °C, investigating every 100 °C, and in 5% H$_2$/N$_2$ (flow rate 25 sccm) from 400 to 1200 °C, investigating every 100 °C, to observe the onset of reaction between GDC and YSZ.

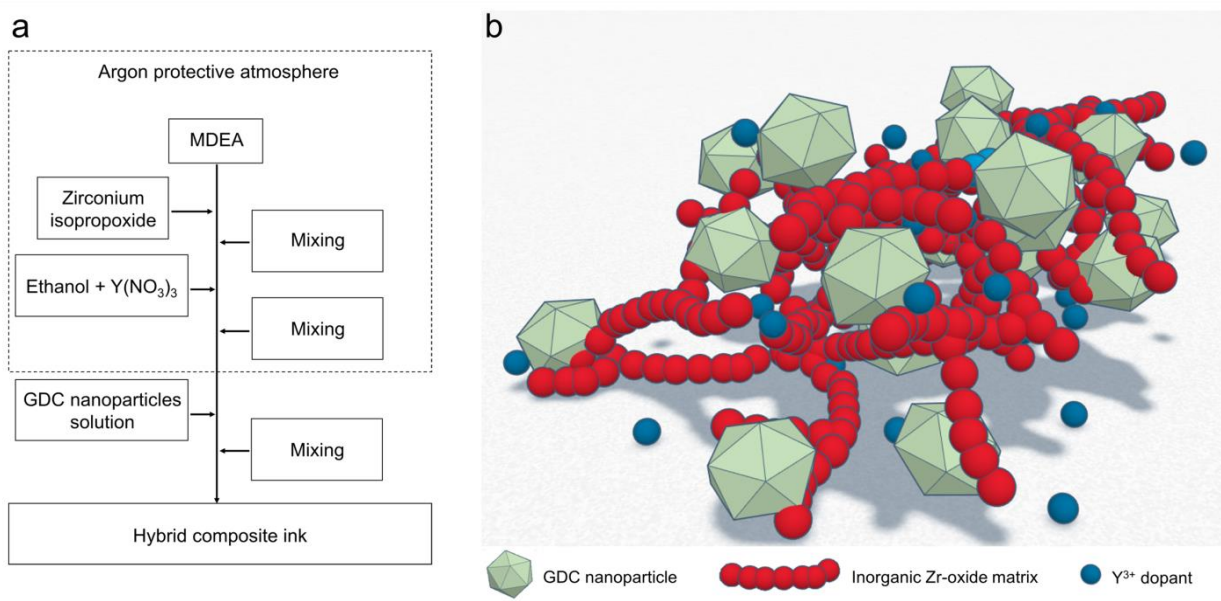

**Figure 1:** Preparation protocol for the hybrid composite ink (a) and schematic representation of the hybrid GDC/YSZ composite electrolyte before the thermal treatment (b).

## 2.2 Samples characterization

The evolution of the crystallographic phase for each material was investigated by X-ray diffraction (XRD) with a Bruker D8 (Cu-Kα radiation, λ=1.54056 Å), using a solid-state detector and acquiring parameters of 0.01 step size, and 0.5 s·step$^{-1}$ time. Diffraction patterns of powders resulting from the ink's calcination at various temperatures in air and 5% $H_2/N_2$ were recorded at room temperature. Both the as-produced particles and the calcined ink were characterized by TEM, using a Jeol JEM 3000F microscope operating at 300 kV.

The SEM characterization of the samples was carried out by a Zeiss MERLIN scanning electron microscope on carbon coated samples.

XPS spectra were carried out with a Perkin Elmer $\phi$5600ci Multi Technique System. The samples were separately fixed by means of conductive carbon tape (Ted Pella) to stainless-steel sample holders. The spectrometer was calibrated by assuming the binding energy (BE) of the Au 4f7/2 line to be 84.0 eV with respect to the Fermi level. Both extended spectra (survey – 187.85 eV pass energy, 0.5 eV·step$^{-1}$, 0.05 s·step$^{-1}$) and detailed spectra (for Ce 3d, Gd 3d, O 1s and C 1s – 23.50 eV pass energy, 0.1 eV·step$^{-1}$, 0.1 s·step$^{-1}$) were collected with a standard Al K$\alpha$ source working at 200 W. The standard deviation in the BE values of the XPS line is 0.10 eV. The peak positions were corrected for the charging effects by considering the C 1s peak at 285.0 eV and evaluating the BE differences.

**2.3 Thin-film fabrication**

In order to perform the electrical characterization of the GDC/YSZ composite, thin films were printed with a Pixdro LP50 inkjet printer. The printer was equipped with DMC disposable piezoelectric printheads from Dimatix that have 16 nozzles with a diameter of 21.5 µm and a nominal droplet volume of 10 pL. Prior to printing, the ink was filtered using a syringe filter with a 450-nm mesh. The thin film was printed on a one-side-polished 1x1 cm square substrate of single crystal (0001) sapphire (Crystal GmbH) using a 400-dpi resolution (corresponding to an estimated printed volume of ca. 0.25 µL/cm$^2$) and subsequently calcined for 6 hours in air at 750 °C, heating and cooling with 1 °C/min ramps. The X-ray diffraction of the film (XRD) was collected in grazing angle geometry ($\omega$=0.5°) with a Bruker D8 (Cu-K$\alpha$ radiation, $\lambda$=1.54056 Å) using a solid-state detector and acquiring parameters of 0.018 step size, and 0.5 s·step$^{-1}$ time.

The SEM characterization of the thin film was carried out by a Zeiss MERLIN scanning electron microscope on carbon coated samples. The density of the film was estimated by analysis of the residual porosity from SEM images, analyzing 20 different micrographs.

**2.4 Electrochemical characterization**

Electrochemical impedance spectroscopy (EIS) was performed in symmetric 2-electrode configuration by using a Solartron 1260 frequency response analyzer over the frequency range of 1 Hz-86 kHz, using 200 mV AC voltages for the acquisition. This configuration is suitable to characterize the in-plane ionic conduction of thin film samples deposited on dielectric substrates, due to large cell constant ($k>10^3$) that allows neglecting the contribution coming from the electrodes to the total resistance[14,42].

To measure the lateral conductivity of the sample, symmetric silver electrodes with ca. 1 mm spacing and 2 mm width were painted on the surface of the thin film. EIS measurements were performed in the temperature range of 400–750 °C in synthetic air (flow rate 50 sccm, waiting at least 1 hour for equilibration before every data acquisition) and the obtained resistance was normalized by the geometrical factors (the thickness of the film was measured after testing by SEM characterization). In order to evaluate the nanocomposite stability, total conductivity measurements were performed at different oxygen activity atmospheres at 600 and 650 °C in synthetic air, nitrogen and 5% $H_2/N_2$ (flow rate 50 sccm). The corresponding oxygen activities have been measured by an in-house built zirconia sensor and determined by Nernst equation.

## 3. Results and discussion

### 3.1 Structural characterization

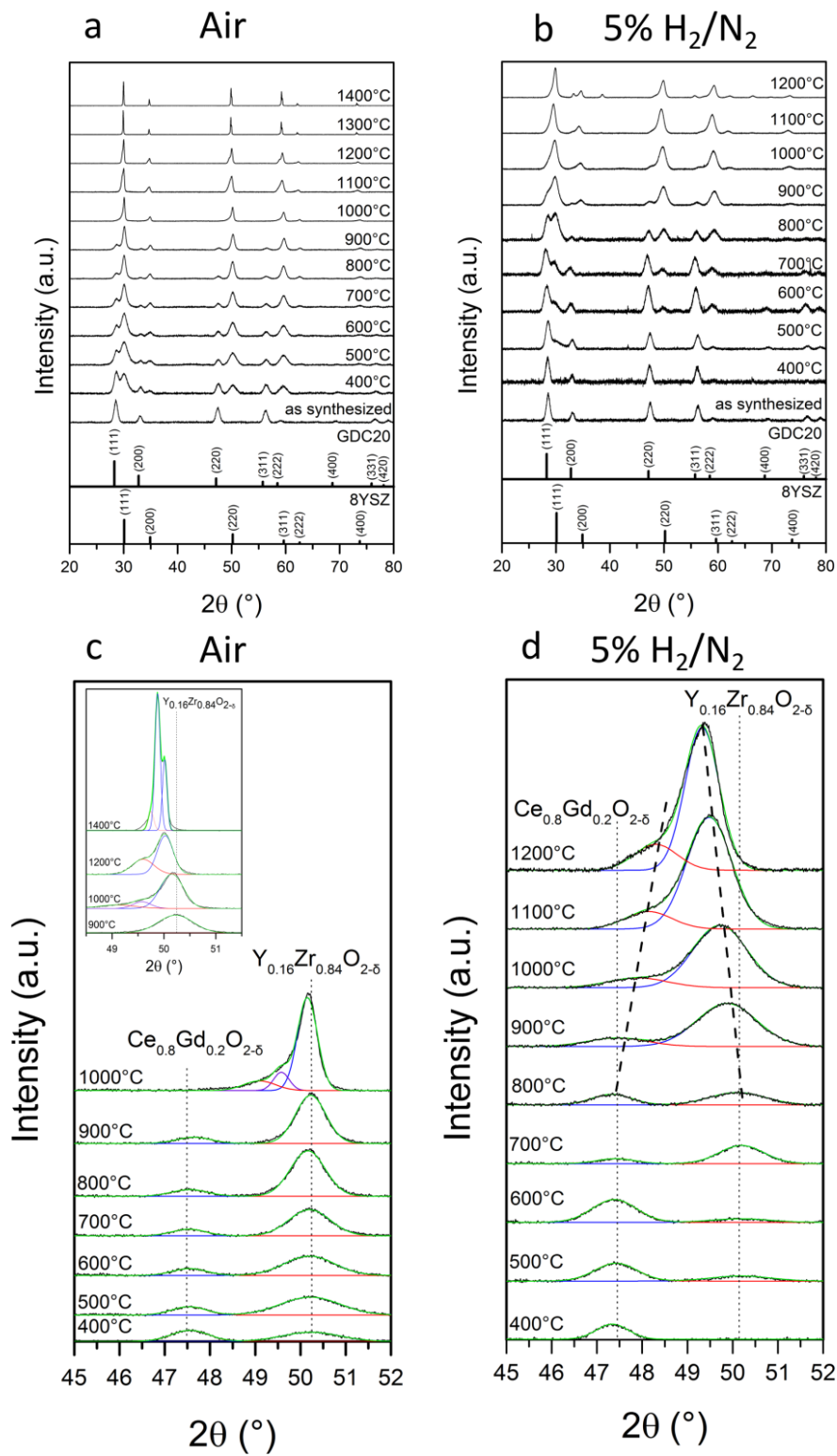

**Figure 2:** Normalized X-ray diffraction patterns of $Gd_{0.2}Ce_{0.8}O_{1.9}/Y_{0.16}Zr_{0.84}O_{1.92}$ (GDC=25% vol/vol) samples calcined in air up to 1400 °C (a) and in 5% $H_2/N_2$ atmosphere up to 1200 °C (b), $Gd_{0.2}Ce_{0.8}O_{1.9}$ (GDC20, JCPDS-ICDD data file card #: 75-0162) and $Y_{0.16}Zr_{0.84}O_{1.92}$ (8YSZ, JCPDS-ICDD data file card #: 30-1468) references are reported for comparison.

Magnification on (220) peak up to 1000 °C of samples calcined in air (c), with inset on high temperature evolution of YSZ peak; magnification on (220) peak up to 1200 °C of samples calcined in 5% $H_2/N_2$ (d).

When interdiffusion occurs, the migration of cations with different valence, ionic radius or coordination induces a modification of lattice parameter in the host lattice. This variation is observable from the shift in the diffraction peaks when compared to those of the original structure. X-ray diffraction is thus able to provide useful information about diffusion processes.

**Fig. 2** shows the evolution of the crystal structure with temperature in air (a, c) and 5% $H_2/N_2$ (b, d). The as-synthesized sample (GDC=25% vol/vol) displays the typical diffraction pattern of cubic $Gd_{0.2}Ce_{0.8}O_{1.9}$ (JCPDS-ICDD data file card #: 75-0162) but no reflections of the $Y_{0.16}Zr_{0.84}O_{1.92}$ (JCPDS-ICDD data file card #: 30-1468). These become clearly visible from 400 °C for the samples calcined in air and slightly evident from 500 °C for the samples calcined in 5% $H_2/N_2$. Compared to our former studies on YSZ produced by sol-gel [38], it is possible to observe that in the case of intimate contact environment with GDC nanoparticles, YSZ formation occurs at 400 °C in air (**Fig. 2a**), a temperature 100 °C lower than what observed in the case pure YSZ. This suggest an influence of GDC nanoparticles on the crystallization of YSZ at lower temperature. It is also possible to notice that removal of the organic framework occurs faster when performed in air (**Figs. 2a** and **c**), with a rapid crystallization of the YSZ phase at low temperature. On the contrary, for the samples calcined in 5% $H_2/N_2$ (**Figs. 2b** and **d**) the formation of YSZ is observed at higher temperature. The influence of atmosphere on crystallization affects the number of grain boundaries, which decrease when grain growth is promoted.

The evolution of the (220) reflections of GDC and YSZ (**Figs. 2c** and **d**) shows that the calcination atmosphere plays a role in the diffusion mechanisms, too. The diffraction data suggest that interdiffusion is activated at a temperature of *ca.* 100 °C lower (900 °C in air and 800 °C in 5% $H_2/N_2$, respectively) compared to what observed in the case of powders having ca. 100 nm size[30]. The large interface between the matrix and the nanoparticles is likely the enhancing factor for the diffusion processes.

Diffusional processes in oxides involve the simultaneous migration of oxygen ions and cations, with the overall process rate being determined by the diffusion rate of the slowest ion. It is well known that in YSZ and GDC the diffusivity of cations is several orders of magnitude lower than the diffusivity of oxygen, with the diffusion of cations represents the limiting step[43–45].

In the case of YSZ, a change in the $pO_2$ is not expected to affect the concentration of defects. For this reason, at a certain temperature, the ionic conductivity is almost constant, independently of

pO$_2$[46]. Similarly, zirconium self-diffusion in YSZ has been observed to proceed through a mechanism based on zirconium vacancies ($V_{Zr}''''$) and not through interstitials ($Zr_i^{\bullet\bullet\bullet\bullet}$) or through the formation of complexes with oxygen vacancies[44], thus being independent on the oxygen partial pressure. Differently from stabilized zirconia, in the case of ceria the oxygen atmosphere can affect the concentration of defects, which is associated with an increase in the concentration of Ce$^{3+}$ at low pO$_2$. Chen *et al.* have observed that, for pure ceria, the lowest energy migration path for cations is represented by octahedral interstitials (Ce$_i$) associated with oxygen vacancies ($V_O^{\bullet\bullet}$)[47]. Thus, the higher the concentration of oxygen defects, the easier the diffusion of Ce$^{3+}$.

Despite being both activated below 1000 °C, the two diffusion processes seem to have different nature: this is expected to be related to the presence of different amounts of oxygen defects in the two investigated atmospheres. In air the diffusion process occurs abruptly, while in reducing conditions it evolves slowly.

Considering the case of air (**Fig. 2c**), the complete disappearance of GDC reflections is observed above 900 °C with the appearance of an asymmetric signal in correspondence to the YSZ reflections, suggesting the onset of formation of a solid solution between GDC and YSZ. Above 1200 °C (inset in **Fig. 2c**), it is observed that YSZ and GDC are not completely mixed to form a solid solution with unique composition, as indicated by the presence of two reflections related to two different cubic phases (**Fig. 2c**, inset). These are detected at lower angles than YSZ (220) reflection, at 1400 °C, suggesting that the stabilization of cubic zirconia due to yttrium doping is in competition with the formation of the ceria-zirconia solid solution. The complete diffusion of zirconium is hindered by the electrostatic effects generated by the trivalent dopants segregating at the grain boundary[48,49].

In the case of reducing conditions diffusion is activated above 800 °C with the progressive shift of GDC and YSZ peaks in opposite directions towards a unique phase composition and appears to be more sluggish. Besides cerium diffusivity being enhanced in reducing conditions (from 10$^{-19}$ cm$^2\cdot$s$^{-1}$ at pO$^2$=0.21 atm to 10$^{-10}$ cm$^2\cdot$s$^{-1}$ at pO$^2$=10$^{-13}$ atm, at 1400 °C)[50], it has been reported before that, for 1% mol doped ceria the grain boundary mobility is severely reduced, since mass diffusion is limited by solute-drag effects[47]. Thus it's likely that the high amount of aliovalent dopant (Gd$^{3+}$) and the cations in YSZ (Y$^{3+}$, Zr$^{4+}$) hinder the diffusivity of cerium through the grain boundary. The possible formation of pyrochlore (Ce$_2$Zr$_2$O$_{7+\delta}$, JCPDS-ICDD data file card #: 08-0221) phases, which could occur at high temperature in reducing conditions, has not been detected, since the (133) and (155) reflections peculiar of this phase have not been observed.

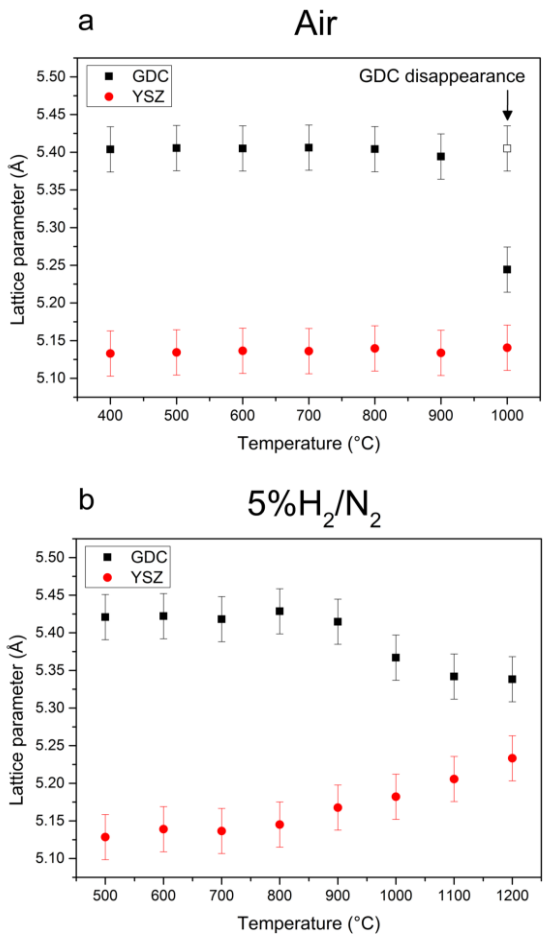

**Figure 3:** Evolution of lattice parameters of GDC and YSZ as a function of temperature, calculated from (220) reflections, for $Gd_{0.2}Ce_{0.8}O_{1.9}/Y_{0.16}Zr_{0.84}O_{1.92}$ (GDC=25% vol/vol) samples calcined in air (a) and in 5% $H_2/N_2$ (b).

In order to facilitate the observation of diffusion processes, a detailed analysis of the evolution of lattice parameters, calculated from the XRD (220) reflections, as a function of temperature is reported in **Fig. 3**. This shows that in reducing atmosphere (**Fig. 3b**) the diffusion processes are activated above 800 °C and their evolution is slower compared to that in air (**Fig. 3a**), where the diffusion is observed above 900 °C. Shifts of GDC and YSZ lattice parameters are detected, with the former observed at a higher angle and the latter at a lower angle. From the calculation of lattice parameters (**Fig. 3b**) it can be noticed that above 800 °C GDC lattice is shrinking, while simultaneously YSZ lattice is expanding. This suggests that, besides $Zr^{4+}$ diffusion towards GDC, which proceeds independently of oxygen vacancies concentration, a migration of $Ce^{3+}$ (r= 1.14 Å in cubic 8-fold coordination) from GDC to YSZ is occurring. The difference in the diffusion rate of $Zr^{4+}$ in the two diffusion atmospheres can be explained considering the different mechanisms involved.

In fact, beside the differences in the ionic radius and valence of the cations, grain boundaries can also play a determining role, too. In air, the abrupt diffusion of $Zr^{4+}$ (r= 0.84 Å in cubic 8-fold coordination) into GDC is favored by $Ce^{4+}$ larger radius (r= 0.97 Å in cubic 8-fold coordination which is 15% larger) and same valence[51]; as an additional factor the large number of grain boundaries, for which Zr diffusivity is reported to be higher than lattice diffusivity ($D_{Zr\_L}\sim10^{-22}$ cm$^2 \cdot$s$^{-1}$, $D_{Zr\_GB}\sim10^{-18}$ cm$^2 \cdot$s$^{-1}$ at 1000 °C and $pO^2$=0.21 atm)[29], is likely to enhance diffusion. Conversely, in reducing conditions $Ce^{3+}$ is formed above 600 °C and this is expected to produce an enhancement of cerium self-diffusion outside GDC lattice, favored by the presence of a high amount of $Ce^{3+}$ (r= 1.14 Å in cubic 8-fold coordination) and $Gd^{3+}$ (r= 1.05 Å in cubic 8-fold coordination) and interstitials[51]. On the other hand, $Zr^{4+}$ diffusion within the GDC lattice is no more favored by having the same valence as the host cation. A difference in charge is reported to create an electric polarization responsible for severely hindering cationic diffusion at the grain boundary by electrostatic repulsion, originating solute drag diffusion effects that can be dramatic at the nanoscale [43,47,52].

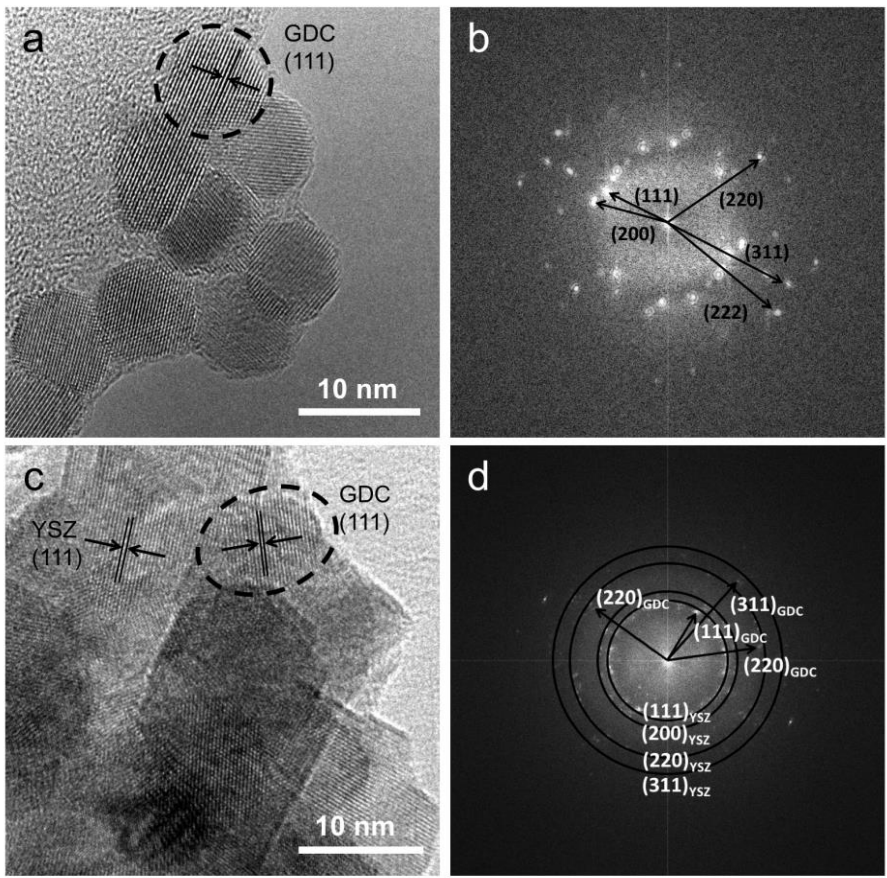

**Figure 4:** TEM images of as-synthesized GDC nanoparticles (a) and relative Fourier transform (b), hybrid $Gd_{0.2}Ce_{0.8}O_{1.9}/Y_{0.16}Zr_{0.84}O_{1.92}$ composite (GDC=25% vol/vol) after calcination in air at 800 °C (c) and relative Fourier transform(d).

**Fig. 4** shows the TEM analysis of the as-synthesized GDC nanoparticles with an average diameter below 10 nm. Fourier transform allows measuring the characteristic (111) lattice spacing of 3.1 Å (b). In presence of YSZ matrix calcined at 800 °C in air (Fig. 3c) sharp grain boundaries are observed, but it is still possible to distinguish round-shaped GDC particles. Their (111) spacing can still be identified from Fourier transform (d), together with YSZ (111) that has a lattice spacing of 2.9 Å. The values obtained for GDC and YSZ lattice spacing are consistent with the values obtained from the XRD characterization.

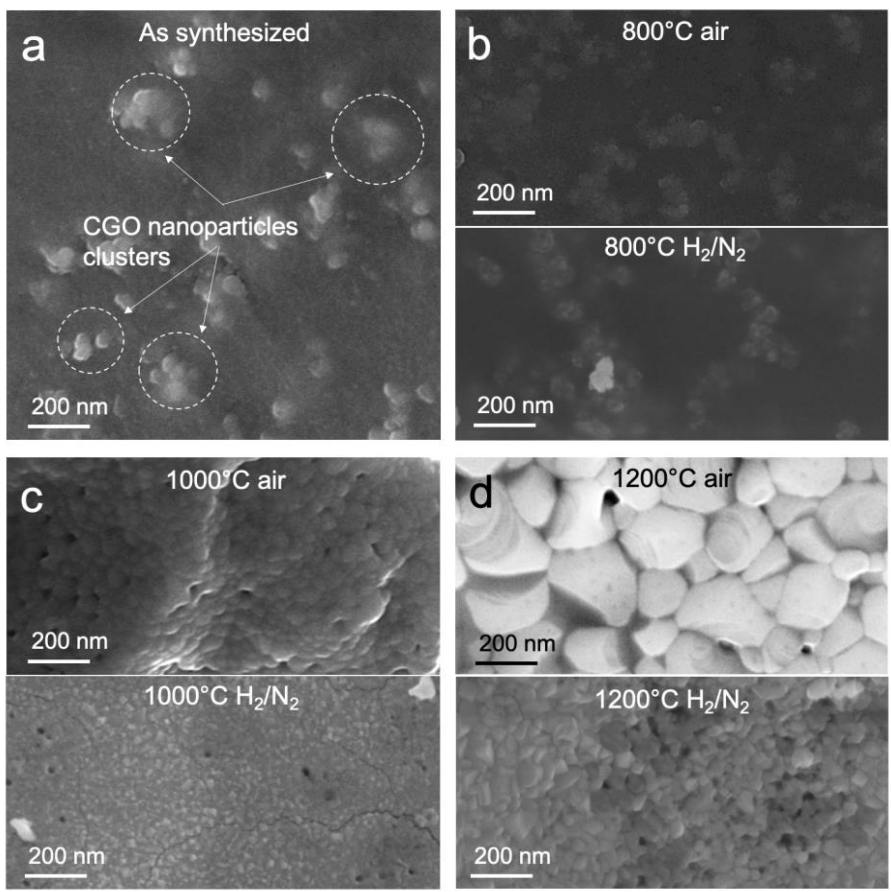

**Figure 5:** Scanning electronic micrographs of the $Gd_{0.2}Ce_{0.8}O_{1.9}/Y_{0.16}Zr_{0.84}O_{1.92}$ (GDC=25% vol/vol) samples: as-synthesized (a) and calcined in air and 5% $H_2/N_2$ at 800 (b), 1000 (c) and 1200 °C (d).

The evolution of microstructure as function of temperature and calcination atmosphere can be followed by SEM. This technique provides complementary information about the interaction between the GDC nanoparticles and the YSZ matrix. **Fig. 5** shows SEM pictures of the microstructure of the samples treated at different temperatures in air and in 5% $H_2/N_2$. In the as-synthesized sample (**Fig. 5a**) a uniform YSZ matrix with no detectable microstructure, in agreement with what was detected by XRD, surrounds small clusters of GDC nanoparticles. The SEM micrographs on the samples treated at 800 °C (**Fig. 5b**) show that the nanoparticle clusters and the surrounding matrix features can still be recognized, in agreement with the results of TEM and X-ray diffraction.

As for the samples treated at 1000 °C (**Fig. 5c**), the calcination treatment in air produces large and rounded crystal grains of *ca* 60 nm that form a homogeneous microstructure in which the features of GDC nanoparticles are not distinguishable anymore. On the contrary, in reducing atmosphere the grain growth is less prominent as a consequence of the reduced diffusion of oxygen and crystal grains of *ca* 10 nm with sharpened grain boundaries are formed. Guo *et al* calculated from EIS measurements an effective grain boundary thickness of *ca* 5 and 7 nm for YSZ[53] and GDC[54], respectively. This suggests that most of the nanocomposite's crystal volume is represented by grain boundary region where solute drag effects can occur. For this reason, despite $Ce^{3+}$ and $Zr^{4+}$ fast self-diffusion in reducing conditions, their interdiffusion results significantly limited at the grain boundaries. The presence of spots with a different electronic density in the micrograph confirms that GDC nanoparticles and YSZ matrix have not formed a solid solution with unique composition.

The microstructure evolution of the nanocomposites at 1200 °C (**Fig. 5d**) shows a growth in the crystal grains for both the samples treated in air and in 5% $H_2/N_2$. The grain dimension of the sample treated in air is *ca* 200 nm, indicating that a remarkable grain growth occurred, leading to densification. This is not observed in the case of the samples treated in reducing environment, in which numerous crystallites with average size below 50 nm and pores are detected.

As a summary of such results we can conclude that, despite being activated at a lower temperature by the low oxygen activity ($10^{-28} < pO_2 < 10^{-14}$ atm between 500 and 1200 °C), the diffusion in reducing conditions results severely hindered. This is a consequence of the slower diffusion of oxygen that does not promote grain growth. Due to that, GDC and YSZ interdiffusion, which is significantly reduced at the grain boundary region by solute drag effect, is limited by grain boundaries, which represent a large volume fraction of nanocrystals and act as bottleneck for diffusion.

To clarify the effect of thermochemical treatment on the materials' interface evolution, XPS spectra of Ce 3d were carried out. Being aware that the valence state of cerium can be severely affected by the measurement conditions, the analysis of every sample should be carried out separately and the acquisition time should be limited to prevent the cerium reduction artefact[55].

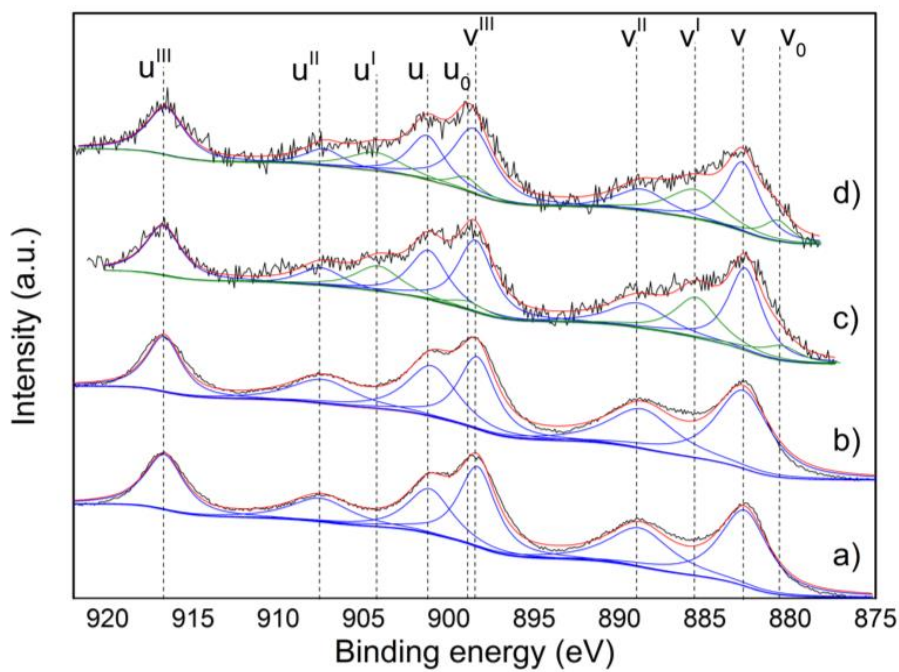

**Figure 6:** Ce 3d photoemission peaks of GDC nanoparticles as-collected (a) and treated at 900 °C in 5%$H_2$/$N_2$ and subsequently re-exposed to air (b) and of hybrid GDC/YSZ (GDC=25% vol/vol) powders from samples calcined at 900 °C in air (c) and in 5% $H_2$/$N_2$ (d).

**Fig. 6** shows the spectrum of cerium with several contributions for each of the 3/2 and 5/2 components. These are marked with *u* and *v* respectively, according to Burroughs' notation[56], and are due to electron correlation phenomena which affect the final state occupation of Ce 4f level[57]. This results in six peaks in the case of $CeO_2$ (*v, v'', v''', u, u'', u'''*) and four peaks for $Ce_2O_3$ (*$v_0$, v', $u_0$, u'*). The two oxidation states have distinct features, so $Ce_2O_3$ and $CeO_2$ pure oxides spectra can be used as references for the interpretation of a mixed valence oxide spectrum. However, when both $Ce^{3+}$ and $Ce^{4+}$ are present, the interpretation of the photoemission spectrum can be complicated by the simultaneous presence of ten peaks. Moreover, a simple linear combination of the pure $CeO_2$ and $Ce_2O_3$ spectra taken as references could not describe the spectrum accurately, since the ratio between the relative intensities of the peaks is suggested not to change linearly with the degree of reduction[58,59].

After Shirley's background subtraction, peaks are fitted, and the weight of the single oxidation state is defined as the sum of peaks areas relative to that oxidation state:

$Ce^{3+} = v_0 + v' + u_0 + u'$

$Ce^{4+} = v + v'' + v''' + u + u'' + u'''$

The concentration of $Ce^{3+}$ can be obtained by the ratio[58]:

$$\%Ce^{3+} = \frac{Ce^{3+}}{Ce^{3+} + Ce^{4+}} \times 100$$

The amount of $Ce^{3+}$ obtained for samples calcined at 900 °C in air (**Fig. 6c**) and 5% $H_2/N_2$ (**Fig. 6d**) is 30% and 34%, respectively. Similar amounts of $Ce^{3+}$ were previously observed by Mullins[59].
No $Ce^{3+}$ signals are observed for the as-collected GDC nanoparticles, with only $Ce^{4+}$ present on the surface of the nanoparticles. For a comparison, the spectrum of the bare GDC nanoparticles treated at 900 °C in 5% $H_2/N_2$ (keeping the gas flow during both the heating and the cooling phase and then exposed to air at room temperature) is reported in **Fig. 6b**. No significant modification is observed, suggesting that GDC nanoparticles undergo a fast re-oxidation. Room-temperature surface oxidation of $Ce^{3+}$ to $Ce^{4+}$ has been observed previously[60].

We can conclude that the presence of the YSZ matrix has an influence on the surface of GDC nanoparticles, suggesting a possible interaction between cerium atoms on the surface and the zirconium precursor occurring during the calcination treatment. Possible reaction paths are suggested:

1) The Zr-containing matrix in which the GDC nanoparticles are completely immersed (as observed from SEM in **Fig. 5**) hinders the interaction of GDC with the surrounding atmosphere during calcination.
2) GDC nanoparticles can take part in the oxidation of the organic framework, losing a part of their oxygen. Cerium oxide is a well-known catalyst for oxidation and is capable of releasing oxygen in a reducing environment and at high temperatures[61].
3) Once the YSZ matrix is formed around the GDC nanoparticles, $Ce^{3+}$ originated at the interface is stabilized by the interaction with the doped zirconium oxide.

It is worth noticing that beside the surface interaction between GDC nanoparticles and YSZ matrix, no significant changes in the GDC crystal structures are detected from XRD in air below 1000 °C, whereas in the samples treated in reducing atmosphere a contraction in the lattice parameter of GDC is detected at lower temperatures (800 - 1200 °C). This effect suggests that in the samples treated in air the presence of $Ce^{3+}$ is limited to the interface with YSZ matrix. On the contrary, a higher amount of $Ce^{3+}$ is present in the samples treated in reducing conditions. The formation of $Ce^{3+}$ at the interface between GDC and YSZ has previously been observed in samples prepared by PLD[62].

**3.2 Thin film characterization**

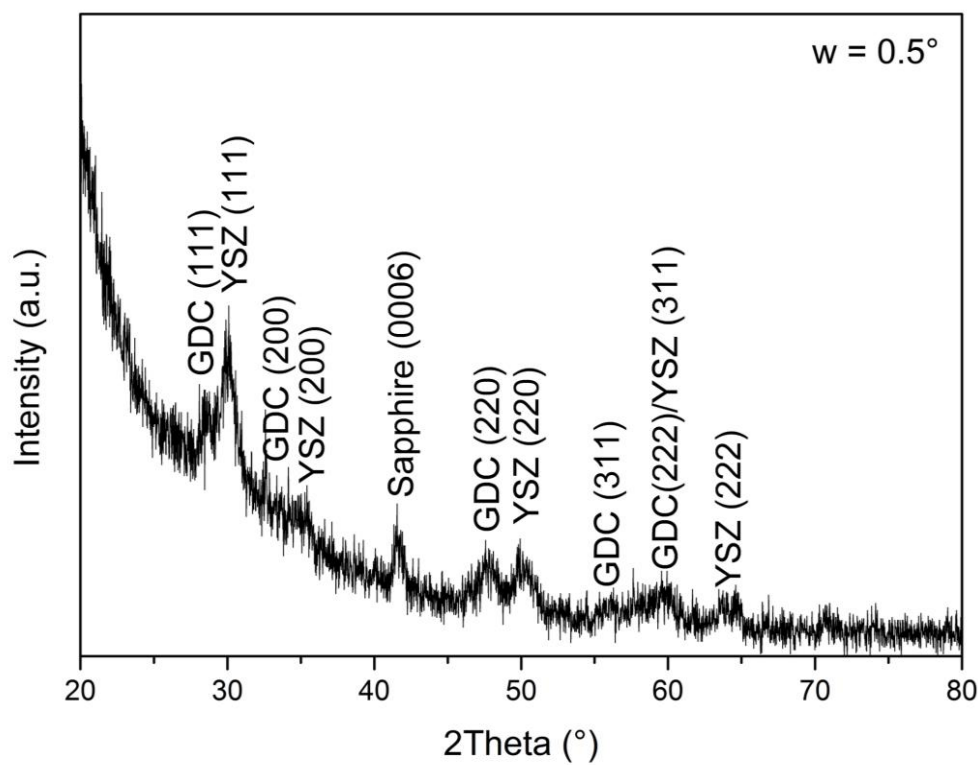

**Figure 7:** Grazing incidence X-ray diffraction pattern of the thin film printed on sapphire (0001) substrate and calcined at 750 °C in air for 6 hours. $Gd_{0.2}Ce_{0.8}O_{1.9}$ (GDC, JCPDS-ICDD data file card #: 75-0162) and $Y_{0.16}Zr_{0.84}O_{1.92}$ (YSZ, JCPDS-ICDD data file card #: 30-1468) references are reported for comparison.

**Figure 7** shows the diffraction pattern, collected in grazing incidence, of the thin film produced by inkjet printing as single layer deposited on a single crystal (0001) sapphire substrate. The reflections of GDC and YSZ are recognizable and no other evident phase except GDC (JCPDS-ICDD data file card #: 75-0162) and YSZ (JCPDS-ICDD data file card #: 30-1468) is observed. The lattice parameter has been determined to be 5.146 ± 0.005 and 5.408 ± 0.005 Å, for YSZ and GDC, respectively. Comparing the lattice parameter with that of reference samples[28], a difference is observed, indicating a 0.3% tensive strain in YSZ and 0.6% compressive strain in GDC.

The average dimensions of the crystallites, obtained from Scherrer's formula, is $5.3 \pm 0.6$ and $8.1 \pm 0.3$ nm for YSZ and GDC, respectively. The dimensions of GDC particles are in agreement with the average size observed before by TEM[40].

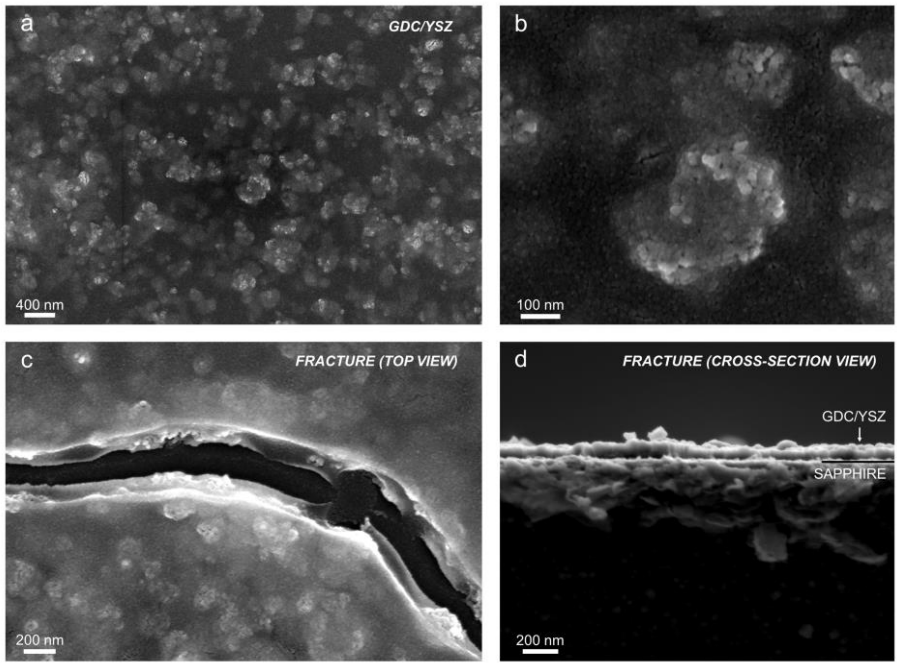

**Figure 8:** SEM images of the thin film printed on sapphire (0001) substrate and calcined at 750 °C in air for 6 hours: top-view pictures showing the dispersed nanoparticles in the matrix forming clusters (a), detail on a cluster of nanoparticles (b), top-view detail on a fracture showing the particles into the YSZ matrix (c); cross section of cold-fractured sample after testing (d).

**Fig. 8** shows SEM observations of the thin film surface and cross section, with the latter being determined after electrochemical characterization. The microscopy (**Fig. 8a**) shows a continuous matrix made of YSZ grains with dimensions below 10 nm in which a high amount of GDC nanoparticles are dispersed in the form of clusters (**Fig. 8b**). This was also observed in the powder-samples (**Fig. 5**), and in the thin film the particles are embedded within the matrix (**Fig. 8c**) which is surrounding them completely. The cross section of the sample (**Fig. 8d**) shows a continuous uniform layer with a thickness below 100 nm. The density of the film was estimated to be $94.4 \pm 2.2\%$ of the theoretical value.

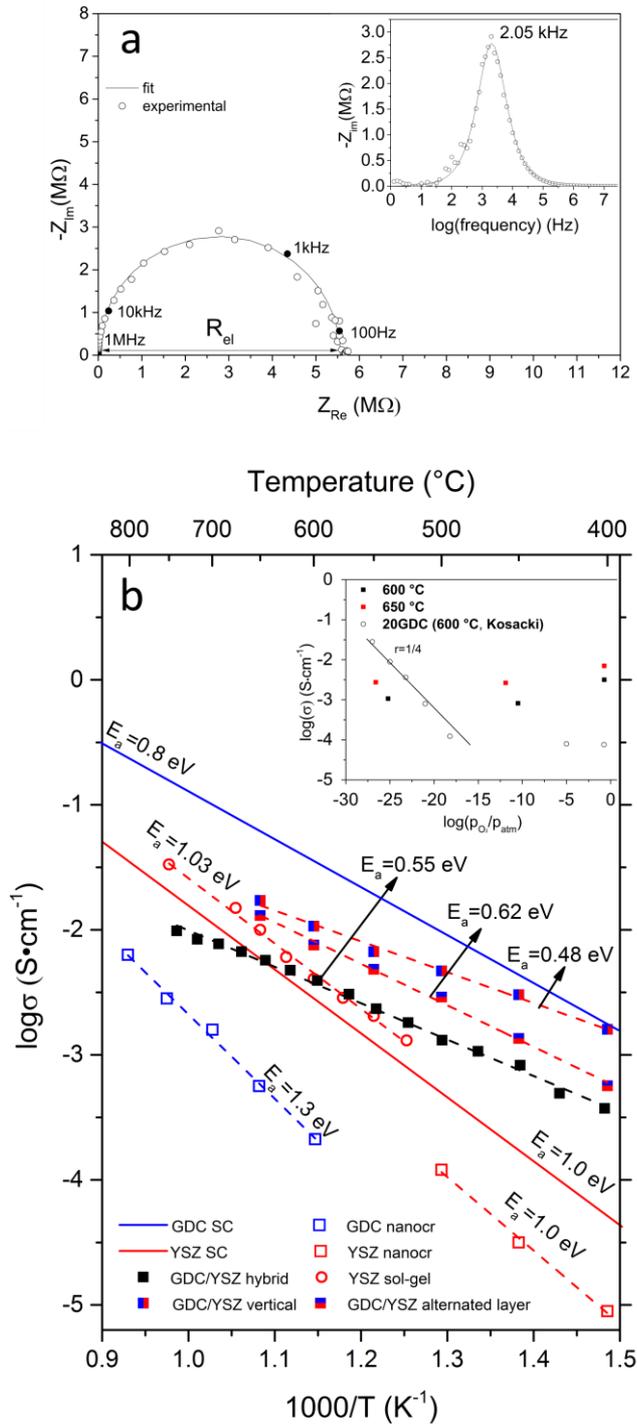

**Figure 9**: Nyquist and Cole-Cole plots measured at 600 °C in air (a), Arrhenius plot of in plane conductivity measured on the GDC/YSZ thin film produced by inkjet printing (b), references values for single crystal GDC (GDC SC) and YSZ (YSZ SC)[63], nanocrystalline GDC (GDC nanocr)[64] and YSZ (YSZ nanocr)[65], YSZ thin film produced by sol-gel inkjet printing (YSZ sol-gel)[38], and bulk YSZ[65] and GDC[66] GDC/YSZ nanocomposites in the form of vertical aligned columns (GDC/YSZ vertical)[67] and alternated layers (GDC/YSZ alternate layer)[68] are reported for a comparison. Total conductivity in function of oxygen partial pressure, measured at 600 and 650 °C in air, nitrogen and 5% $H_2/N_2$ is reported as inset with Kosacki values for 20GDC ($d_G$=36 nm)[64].

**Fig. 9** summarizes the results of the conductivity measurements. The Nyquist plot (**Fig. 9a**) shows a characteristic single semi-arc, well fitted by a single R//C Voigt element, at all temperatures in which grain boundary and bulk contributions are not separated[69]. The Cole-Cole plot of the same data is also included as an inset.

This result is typical of two electrodes in-plane testing configuration and allows neglecting contribution of electrodes to the total conductivity, due to the large geometrical factor[69]. However, the capacitance (~$10^{-11}$ F) obtained with this setup must not be confused with a grain interior contribution: in fact, it is the result of the contributions of all involved capacitances, including grain interior, grain boundary and stray capacitance. The stray capacitance arises mainly from the substrate and is of complex determination, typically making impossible to separate the contributions from grain interior and grain boundary.

The overall characterization indicates an ohmic resistance that is measured at the relaxation frequency of 10 Hz, obtaining a value of 5.6 M$\Omega$, and the apex frequency for the capacitive effect at 2.05 kHz at 600 °C. The resistance values obtained at different temperatures in air are normalized by the geometrical factors and the total conductivity of the sample is plotted as Arrhenius plot in **Fig. 9b**.

The Arrhenius plot of the conductivity of GDC/YSZ thin film measured in air includes reference values for single crystal and nanocrystalline YSZ and GDC[9]. The conductivities of two nanocomposites with different geometries, columnar and planar, are reported[67,68]. Values measured for the YSZ thin film produced with sol-gel reactive ink are also included[38].

It can be observed that the nanocomposite has a higher conductivity than nanocrystalline GDC and YSZ, which are typically characterized by a lower conductivity due to the blocking effect of grain boundaries.

The conductivity values observed at low temperatures show a beneficial effect of GDC presence in the composite at low temperatures, leading to a higher conductivity than singe crystal and thin-film YSZ below 600 °C; however, the thin film obtained by inkjet printing is characterized by a lower conductivity than other nanocomposites prepared by pulsed laser deposition (PLD)[67,68]. A possible reason for this difference is that the highly ordered structures obtained by PLD typically come from pre-sintered targets which have been treated at high temperature and thus have intrinsically a higher degree of crystallinity. Since the microstructure of the film obtained by inkjet printing is characterized by a low porosity, a difference in the interface between GDC and YSZ has been ruled out. In fact the activation energy determined for our film (0.55 eV) is comparable to the values obtained by Azad *et al* (0.62 eV) and by Su *et al* (0.48 eV) in the case of alternated layers and vertically aligned columns composites, respectively[67,68].

As previously observed, the coupling of nanometric GDC and YSZ is characterized by the presence of fast contributions coming from the interfaces of the composite, where the microstrain creates lower energy migration paths for oxygen vacancies[22]. This is confirmed by the differences in the lattice parameters detected from GI-XRD, which indicate a 0.3% tensile stress in YSZ, which has been reported to enhance the ionic conductivity[67]. We can't exclude that the presence of $Ce^{3+}$ observed at the interface between the nanoparticles and the matrix could be beneficial too, since it would be compensated by a higher amount of oxygen vacancies[62].

Since the presence of $Ce^{3+}$ at the interface has been observed by XPS, the total conductivity is measured at different oxygen partial pressures (**Fig. 9b**, inset) in order to exclude possible electronic conduction contributions coming from GDC. Ceria based electrolytes typically display an abrupt enhancement of conductivity in reducing environments, which is associated with the partial reduction of cerium oxide at low oxygen activities ($pO_2 < 10^{-25}$ atm) and the onset of n-type electronic conductivity[64]. This behavior is well known in ceria-based materials and is due to the formation of $Ce^{4+/3+}$ couple responsible for electronic conduction, which is promoted in presence of a high amount of dopants[70].

From the total conductivity measurements (**Fig. 9b** inset), it can be noticed that the total conductivity of the nanocomposite is not significantly changed in reducing environment, while in the case of GDC20 with 36 nm large grains an increase of the total conductivity of several orders of magnitude due to n-type conduction has been observed [64]. This suggests that the chosen amount and the uniform dispersion of GDC nanoparticles does not produce a path of percolation for the electronic conduction, ensuring constant electrolyte performances over a wide range of $pO_2$.

## 4. Conclusions

A novel $Gd_{0.2}Ce_{0.8}O_{1.9}/Y_{0.16}Zr_{0.84}O_{1.92}$ nanocomposite can be produced as an ink-jet printable thin-film by a novel hybrid chemical method consisting in uniformly dispersing of GDC nanoparticles into a YSZ sol-gel reactive matrix. This approach allows obtaining an inkjet-printable composite material suitable for low-temperature processing.

The phase and microstructure evolutions of the composite show both in air and reducing environment chemical and structural stability with no detectable signs of ceria chemical reduction.

No interaction between the two materials is observed below 1000 °C in air, when a rapid diffusion of zirconium towards GDC occurs. On the contrary, in reducing environment, despite the diffusion of $Ce^{3+}$ being activated at lower temperatures (900 °C), the interdiffusion process is characterized by slower evolution, hindered by the large amount of grain boundaries.

The nanocomposite electrolyte is deposited by ink-jet printing and calcined at low temperature, obtaining a thin film with uniform thickness below 100 nm in which the GDC nanoparticles are embedded in the YSZ matrix, inducing a microstrain in YSZ. The resulting hybrid GDC/YSZ electrolyte displays a synergic coupling of the YSZ matrix thermochemical resistance with an elevated ionic conductivity enhanced by the presence of non-percolating GDC nanoparticles. The high ionic conductivity at low temperatures appears to be dominated by interfacial effects between the oxide components.

**Conflicts of interest**

The authors have no conflicts to declare.

**Acknowledgements**

This project has partially received funding from the Fuel Cells and Hydrogen 2 Joint Undertaking under grant agreement No 700266. This Joint Undertaking receives support from the European Union's Horizon 2020 research and innovation program and Hydrogen Europe and N.ERGHY. Giovanni Perin gratefully thanks Fondazione Cariparo for the financial support.